\pdfoutput=1

\documentclass[12pt,letterpaper]{article}

\usepackage[totalwidth=440pt, totalheight=640pt]{geometry}
\usepackage[T1]{fontenc}
\usepackage[utf8]{inputenc}
\usepackage{lmodern}
\usepackage{amsmath}
\usepackage[square,numbers]{natbib}
\usepackage{url}
\usepackage{hyperref}
\makeatletter
\newcommand{\address}[1]{\gdef\@address{#1}}
\newcommand{\email}[1]{\gdef\@email{\url{#1}}}
\newcommand{\@endstuff}{\par\vspace{\baselineskip}\noindent\small
	\begin{tabular}{@{}l}\scshape\@address\\\textit{E-mail:} \@email\end{tabular}}
\AtEndDocument{\@endstuff}
\makeatother
\usepackage{subcaption,graphicx}
\usepackage{color}
\usepackage{float}
\usepackage{bm}
\begin{document}

\newcommand{\macro}{\newcommand}	
\macro{\be}{\begin{equation}}
\macro{\ee}{\end{equation}}
\macro{\eps}{\epsilon}
\macro{\etal}{{\em et al }}

\title{Lattice approach to plane colorings}
\date{}
\author{Sami Heinäsmäki}
\address{Aalto University, P.O.Box 11000, 00076 AALTO, Finland}
\email{sami.heinasmaki@aalto.fi}
\maketitle

\begin{abstract}
	I propose a fixed-range interaction multicomponent spin model, to be used as a physical analog to problems in plane geometry. Specifically, the model is applied to the open problem of the chromatic number of the plane. When spin values are interpreted as colors, the lowest energy configurations of the lattice spin system can be interpreted as approximations to plane colorings. In general minimum energy configurations of the model give optimal colorings, corresponding to minimum probability of any color realizing distance one. Approximate optimal lattice colorings with two to seven colors towards the continuum limit suggest that a true coloring of the plane cannot be achieved with less than seven colors.   
\end{abstract}

\section{Introduction}
Imagine throwing a stick of unit length randomly on an infinite Euclidean plane, colored with $q$ colors. If the endpoints of the stick always fall on different colors, the plane is said to exhibit a $q$--\textit{coloring}\footnote{In other words: none of the colors realizes distance 1.}. The smallest such number $q$ is the \textit{chromatic number of the plane}.

The chromatic number of the plane is currently known to be either 5, 6, or 7. The upper bound is based on explicit colorings of the plane, satisfying the unit-distance condition. The exclusion of the lower values 2--4 can be shown using special unit distance graphs. In the simplest case, consider an equilateral triangle with side length 1 on a plane. Two of the vertices are necessarily on a same color, thus proving a two--coloring impossible. Likewise, a slightly more complicated seven--vertex \textit{Moser spindle}~\cite{moser} shows a three--coloring impossible. This was the status of the question, also called the Hadwiger--Nelson problem, from its inception in the 1950's, until de Grey in 2018 constructed a 1581--vertex graph~\cite{degrey}, which proved the impossibility of a four--coloring\footnote{The graph has since been reduced to 553 vertices~\cite{polywiki}.}. The history of the problem prior to de Grey's finding is described in great detail in the book by Soifer~\cite{soifer}, which includes discussions of known results and related problems.

In this note I propose a method which does not directly aim at a proof, but rather provides an alternative view to the problem. It is an experimental approach, inspired by the statistical stick--throwing argument above. The method is based on explicit, but necessarily approximate, attempted colorings of the plane. They are obtained as lowest energy configurations of a spin model, when the spin values are interpreted as colors. The model is defined on a lattice, and the colorings in a mathematical sense correspond to the continuum limit of the model. The behavior of the computed ground states as the lattice approaches continuum is taken to indicate whether the corresponding plane coloring is possible. 

The model proposed here also gives rise to a natural generalization: one may ask not only when none of the colors realizes distance one, but when the probability to find equal colors separated by unit distance attains a minimum value for a given number of colors, even when zero value is not possible. The minimum energy principle automatically points to such \textit{optimal colorings} as ground states of the model. In the graph--theoretic setting a probabilistic version of the problem has been stated by Bourgeat~\etal~\cite{proba}, where the fraction of bi--chromatic sections of the unit graph was set to be maximized. This definition results in an identical criterion as the one used here. Moreover, their paper contains important analytical results, which can be used to evaluate the performance of the current method.

The numerical approach comes with many caveats, discussed below, but to the extent computations have been performed here, they point to some general features. The computed optimal results for two to four colors are map--type colorings, and especially the two and three color cases are in agreement with the analytical results in~\cite{proba}. For five and six colors the optimal results are not purely map--type but contain diffuse areas. Whether the plane can be colored with five or six colors is not resolved here, but only in the case of seven colors there were examples in the present simulations that converged to zero probability in finding equal colors separated by unit distance.

Another argument for the existence of colorings can be based on \textit{frustrations} in the lattice, defined as locations where spins have high energy. As discussed below, these correspond to same--colored sites unit--distance apart within the proposed model. The best--converged lattice colorings for five and six colors showed periodic structures of frustrations which are unlike numerical noise and whose existence indicates that there are obstacles for coloring the lattice with less than seven colors. If the trend persists in further analysis, it suggests that the chromatic number of the plane is seven.

Spin models based on long--range interactions are not commonly met, but there is at least one previous study~\cite{goulko}, where Goulko and Kent used a two--spin model with a constant total magnetization and fixed interaction length to study another problem in plane geometry. Their paper is recommended reading, as it provides more background on the numerical analysis of these types of systems. 

\section{The model}
The model proposed here is technically a unit--range interaction Potts model, with integer-valued spins ${\bm s}_i$ ($1 \leq {\bm s}_i \leq q$) located at lattice sites ${\bm r}_i$. The lattice structure does not naturally allow specifying a fixed range of interaction, but it can be approximately enforced by employing an auxiliary one--dimensional distribution function $\phi$ with unit integral. When $a$ is the lattice constant (also called lattice spacing), the model is described by the Hamiltonian
\be
H = \sum_{ij} \delta_{{\bm s}_i {\bm s}_j} \, a^{-1}
\, \phi \left( \frac{|{\bm r}_i - {\bm r}_j| - 1}{a} \right)
\label{h1}
\ee
where the summation extends over all pairs of lattice sites. The distribution $\phi$ is assumed to satisfy also certain additional conditions to mitigate the effects of the discrete lattice structure on the computations, discussed in more detail by Peskin~\cite{peskin}. These conditions include e.g.~bounded support for computational efficiency, but besides those restrictions $\phi$ can be left unspecified for now. The family of Hamiltonians~\eqref{h1} has a formal unique continuum limit, which follows from the limiting behavior of distributions 
\be
a^{-1} \phi \left( \frac{|{\bm r}_i - {\bm r}_j| - 1}{a} \right)
\xrightarrow[a \to 0]{}  \delta(|{\bm r}_i - {\bm r}_j| - 1).
\label{philim}
\ee
 This immediately points to the main idea: the energy (the value of the classical Hamiltonian for a given configuration of the spins) is lowest when the number of equal spins located at lattice sites separated by unit distance is minimized. If the energy is zero, all the spins separated by unit distance are different, and when the individual spin values are interpreted as colors, we have a coloring of the lattice. The main hypothesis here is that one may search for the smallest number of spin values realizing the zero energy in the continuum limit, and this should equal the chromatic number of the plane.

A suitably normalized lattice constant--dependent energy functional can be defined by taking the ratio of the energies of the current spin configuration and one where all spins are equal. It is the energy value targeted in the simulations, and can be computed by taking the average of the corresponding single--site quantities: 
\be
   \eps(a) = \frac{1}{N} \sum_{k=1}^N \eps_k (a)
   \label{ene}
\ee
where $N$ is the number of lattice sites and
\be 
  \eps_k (a) = \frac{1}{D_\phi (a)} \sum_{i=1}^N \delta_{{\bm s}_i {\bm s}_k} \phi \left( \frac{|{\bm r}_i - {\bm r}_k| - 1}{a} \right).
  \label{epsk}
\ee
The normalization factor $D_\phi (a)$ is the maximum energy of a single site. It is a constant factor for each choice of the distribution $\phi$ and the lattice constant $a$:
\be
   D_\phi (a) = \sum_{i=1}^N \phi \left( \frac{|{\bm r}_i - {\bm r}_j| - 1}{a} \right)
\ee
which is independent of the reference site $j$. A common factor of $a^{-1}$ has been omitted above. The functional $\eps(a)$ stays finite in the limit of an infinite lattice $N \to \infty$ and further when the continuum is approached in the additional limit $a \to 0$ (in the simulations below $Na^2$ was kept fixed, but of course this merely tends towards a finite region of the plane). 

It follows from~\eqref{ene} that always $0 \leq \eps (a) \leq 1$ and thus one can also view $\eps(a)$ as the probability of finding equal spins separated by unit distance in the lattice. This is utilized below, when I search for \textit{optimal colorings} in the cases $2 \leq q \leq 7$, defined by $\eps(a)$ attaining a minimum value. 

Besides the overall lattice average $\eps(a)$ in~\eqref{ene}, one can in addition look at the distribution of the single--site relative energies $\eps_k (a)$~\eqref{epsk} over the lattice. A regular distribution of nonzero values is suggestive of a converged local minimum in the spin system. This can be contrasted to an irregular distribution, which is natural to identify with numerical noise. The distributions of the values $\eps_k (a)$ can be analyzed together with the convergence of the whole--lattice average $\eps(a)$, and they are useful in visualizing the results of the lattice computations below.

Note that there exists an alternative concept in the literature, which can be used as a criterion of goodness of a given coloring~\cite{soifer2,soifer}. This is the \textit{coloring type}, which is a coloring of a plane with $n$ colors and denoted by $(d_1,d_2,\ldots,d_n)$ such that color $i$ ($1 \leq i \leq n$) does not realize distance $d_i$. The goodness is then measured by how many of the $d_i$ can be made equal to one. While this concept is useful when explicit colorings are constructed, I use the probabilistic measure here, as it is natural in the context of the present spin model analogy where all colors are treated on equal footing.

Among the properties of the Hamiltonian~\eqref{h1} is translational invariance in the lattice, and this should give rise to colorings which in the continuum limit have translational invariance at long distances (appreciably longer than unit distance). Also, the ground state of the model cannot be unique in the continuum limit for $q \geq 7$, due to the existence of several explicit seven colorings of the plane~\cite{soifer}. Whether the ground state is degenerate for lower $q$ values also can affect the efficiency of the search of the lowest energy configuration, as it happens (in this work) locally, and could in principle converge towards different minima in different regions of the lattice.

 From the implementation point of view the essential feature of the model is that the approximation for the fixed--range interaction is dependent on the lattice constant $a$. This forms the basis for the study of the model's behavior when approaching the continuum limit. The bounded support of distributions $\phi$ restricts the number of interacting neighbors each lattice site has, but this number is still large and grows towards the continuum limit, in stark contrast to spin models with nearest--neighbor interactions. As an example, for a lattice constant $a = 0.1$ each site has 240 interacting neighbors, whereas for $a = 0.01$ the number is 2520, provided $\phi$ satisfies the properties discussed in~\cite{peskin}.

The lattice serves as an analog to the plane, but an additional approximation is necessary in simulations: the lattice must of course be finite. This potentially causes finite--size and boundary effects, which are collectively called \textit{volume effects}. As seen later, these were in in some forms present in the results, but the severity of the volume effects should however not be fatal due to the translational invariance of the Hamiltonian~\eqref{h1}. The long--distance invariance is indeed approximately visible in the results shown later.

\section{Lattice computations}
Six different spin systems were studied in several sets of computations. The number of spin values (referred also as colors in what follows) varied from 2 to 7. A default set of computations used a lattice of $10 \times 10$ unit areas, with four different lattice constants 0.1, 0.05, 0.025, and 0.0125. Also additional computations with larger and smaller lattices were performed, as were computations with smaller lattice constants. This was done to analyze the convergence properties of the computations as a function of $a$, and to estimate the volume effects on the generated lattice colorings.

The search for the lowest energy configuration, measured by the quantity $\eps(a)$ in~\eqref{ene}, was performed by simulated annealing. In this approach one starts from a random configuration of spins, and performs a series of Monte Carlo updates of the lattice. The probability to accept updates increasing the individual site energies is gradually lowered (effectively, lowering the temperature of the system), until the spins approach a fixed configuration. The Metropolis algorithm was used for updating the spins. This amounts to generating a new candidate spin for the lattice site, computing the energies $\eps_{new}$ and $\eps_{old}$ according to~\eqref{epsk}, and finally updating the spin with probability ${\rm min} \left[1, \exp((\eps_{old}- \eps_{new})/T)\right]$. Here $T$ is an effective temperature--like quantity, whose initial value is proportional to the site energy in a random spin system. An essential part of simulated annealing is the scheme to lower $T$ in the course of simulation. In this work a simple scaling was employed, where a new $T$ value was computed as $T_{i+1} = cT_i$, using values $0.83 \leq c \leq 0.95$. For each $T$ value the lattice was repeatedly updated so that $\eps(a)$ converged to a roughly constant value, before lowering $T$. In all computations a square lattice was used, and the spins were updated with typewriter ordering. There is no canonical way to perform the simulated annealing process, and I encourage interested readers to write their own simulators and test different schemes. The simulation code used in this work is available as free software~\cite{cp}, including the input files used to generate the results presented here\footnote{Exact reproduction of the results would require a known initial state of the pseudorandom number generator. This is not utilized here, but the lattices are available for download.}.

Choosing the boundary conditions of the lattice is essential in the spin update process. Here free boundaries were used, meaning that the lattice spin values $1 \cdots q$ were treated as being surrounded by zero spins, which were never updated. They thus didn't give contribution to the energy. The zero value played the role of a constant extra color. This was judged to have the least effect on the lattice spins. Boundary effects could be minimized by cyclic boundary conditions, but this would change the underlying topology, and the model would not be a plane analog anymore. 

The following distribution was used here for approximating the lattice delta function:
\be
\phi (x) = \left\{ \begin{array}{ll}  \frac{1}{4} 
	\left( 1 + \cos \left( \frac{\pi x}{2} \right) \right),
	& |x| \leq 2 \\ 0, & |x| > 2  	\end{array}  \right.
\label{phi}
\ee
Other possible forms are discussed e.g.~by Yang~\etal~\cite{yang}, but were not investigated here. Discretization effects related to the choice of the distribution $\phi$ in lattice computations were also studied by Goulko and Kent~\cite{goulko}, who used~\eqref{phi} as a default distribution. Notwithstanding the particular approximation used, the most important factor affecting $\eps(a)$ is that the finiteness of the lattice makes the computed $\eps(a)$ always smaller than the "true" value. This follows from the long range of the interaction: for a lattice size of $L \times L$ unit areas, only a fraction of $(1 - 2/L)^2$ of the sites have all interacting neighbors inside the lattice, the rest being "border spins" having less neighbors and thus contributing less to $\eps(a)$ (remember that all individual contributions to $\eps(a)$ are always positive). For $L=10$ this fraction is 0.64 and for $L=6$ it is only 0.44. Nevertheless, as seen later, the effects resulting from this are not dramatic and can be analyzed by computing $\eps(a)$ cumulatively for the lattice, starting from the interior. The fact that lowering $a$ increases the number of interacting neighbors each lattice site has, is the most important limiting factor in the computations. As a rough rule, halving $a$ (while keeping $Na^2$ constant) increases the computation time tenfold.

A summary of the computed energies is given here, and a more comprehensive discussion of the computations and the associated lattices is given in the following sections. A collection of best estimates for the energies $\eps(a)$ is given in Table~\ref{epsvals}. The energy values $\eps(a)$ were approximately corrected to take into account the finite size of the lattice using a method described in the end of Sec.~\ref{kakkonen}, but they still represent rather crude estimates. For analytical bounds in cases $q=2,3,4$, see~\cite{proba} and~\cite{polymath2}. Besides the values $\eps(a)$ for the six cases studied, the area of the lattice consisting of sites which have no same--colored neighbors unit distance away is given in percentages. This ratio is computed by taking the number of $\eps_k (a)$ values equaling zero and dividing that by the total number of lattice sites. It should be noted that this number is not corrected to take into account the finite size of the lattice, and thus overestimates the true value. The table includes also the lattice parameters employed in the corresponding simulations.

The results at this level display a rough trend where the probability to find same--color neighbors drops by an order of magnitude when the number of colors is increased by one. The simulation result for the two spin case is the correct global minimum, and it means that the solution of the present model is known exactly for two spins in the continuum limit. Also the computed optimal four--color lattice was relatively simple, and it should be possible to solve the energy of this particular configuration exactly. The result for seven colors was zero, in agreement with the known existence of seven colorings of the plane.
\begin{table}
	\centering
	\begin{tabular}{clcccc}
		\hline \hline
		Colors & $\eps(a)$ & Colored area (\%) & Lattice area & $a$ \\ \hline
		2 & 1/3 & 0 & $20 \times 20$ & 0.0125  \\ 
		3 & $0.12$ & 17 & $10 \times 10$ & 0.0125 \\ 
		4 & $0.01$ & 72 & $10 \times 10$ & 0.0125 \\
		5 & $0.004$ & 81 & $10 \times 10$ & 0.0125 \\ 
		6 & $0.0002$ & 97 & $6 \times 6$ & 0.0060 \\
		7 & 0.0 & 100 & $6 \times 6$ & 0.0125 \\
		\hline \hline
	\end{tabular}
	\caption{\label{epsvals}Numerical estimates for the optimal values for energies $\eps(a)$ in~\eqref{ene}, the area of the lattice where the sites have no same--colored neighbors unit distance away (in percentages), and the associated lattice parameters used in those calculations. For two colors an analytical result for the simulated configuration is referred, and the computed result for seven colors was zero within the double precision arithmetic, but is known to be exactly zero in the continuum limit. In other cases $\eps(a)$ values are rough estimates obtained using a method described in the end of Sec.~\ref{kakkonen}.}
\end{table}

\begin{figure}[t!]
	\centering
	\includegraphics[width=0.95\linewidth]{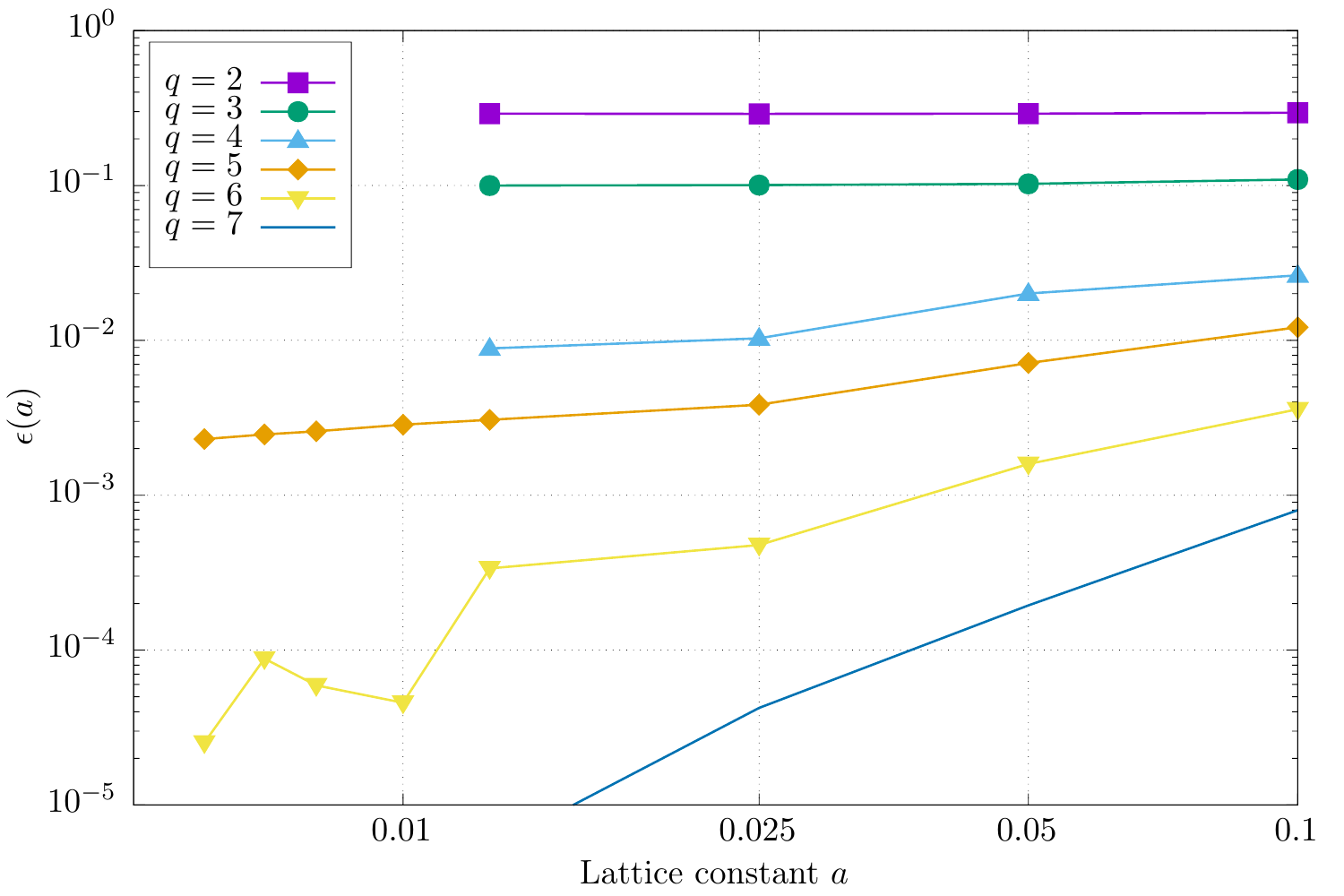}
	\caption{\label{cvg}Convergence of the parameter $\eps(a)$ in~\eqref{ene} as a function of the lattice constant. Note the logarithmic scales on axes. The case $q=7$ was the only one which converged to zero (not shown in the log--scale plot). In cases $q=5,6$ there was apparent convergence towards small but finite values, although this could not be resolved definitely. The given values are not corrected against the finite size of the lattice, but estimates for the final $\eps(a)$ values are given in Table~\ref{epsvals}.}
\end{figure}

The convergences of the energies $\eps(a)$ towards the continuum limit are shown in Fig.~\ref{cvg}. For all numbers of colors below seven the convergence slowed or halted for small $a$, whereas the seven color case converged to zero. While only a limited set of computations was performed in this work, and very small lattice constants were not used in simulations due to the associated computational cost, the computed values were rather consistent across simulations. The values given in Fig.~\ref{cvg} were obtained in computations using several different lattice sizes, and they are not corrected for finite--size effects, which are very small at this level of visualization.
The evolution of lattices in the case $q=2$, corresponding to each lattice constant value, is discussed in more detail in Sec.~\ref{kakkonen}. In the cases $q = 3 \cdots 7$ only one example lattice for each number of colors is shown in the sections below, although more lattices are available for download along with the code~\cite{cp}.

\subsection{\label{kakkonen}Two colors}
There is no true two--coloring of the plane, but performing the annealing procedure for two (as well as three and four below) colors gives important feedback of the performance of the proposed scheme. For the model to be a useful analog to the original problem, $\eps(a)$ should converge to a finite value at least for all $q < 5$, when $a$ tends towards zero. An important test here is comparison with the exact solution found in~\cite{proba}, which states that  the optimal two--coloring consists of alternatingly colored stripes with width $\sqrt{3}/2$, corresponding to probability $1/3$ for finding equal colors unit distance apart.

I discuss the two--color case in more detail, since here it is possible to compare the numerical results to solvable reference configurations. One can estimate the goodness of the solutions, and visualize some general features characteristic to the current approach. It turns out that the the numerical results point to some general potential problems related to finite lattices, but also, that the optimal two--coloring is indeed achieved by this method. This gives hope that the model could find the optimal coloring of the plane with higher number of colors also.  

\begin{figure}[h!]
	\centering
	\includegraphics[width=0.8\linewidth]{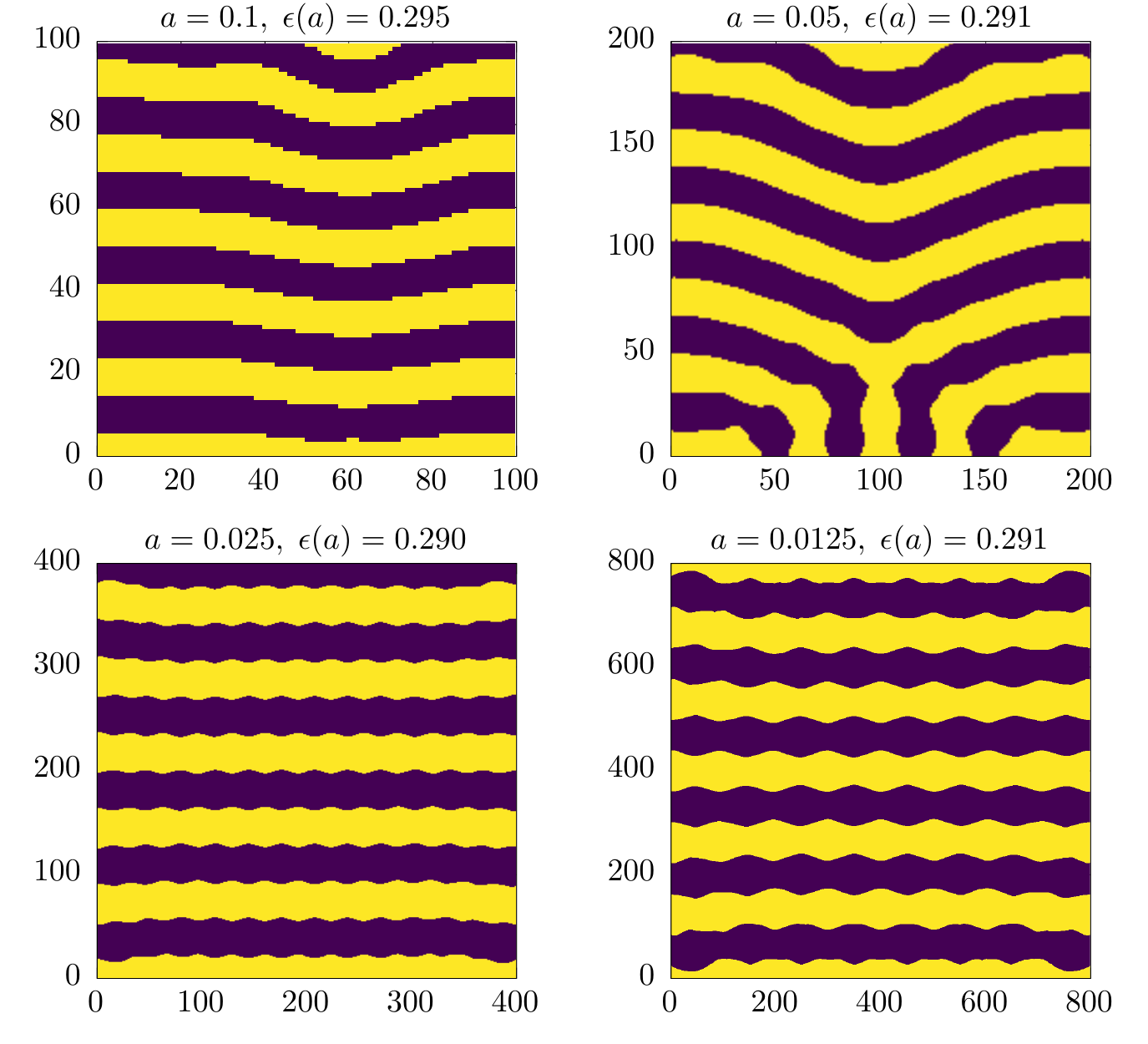}
	\caption{\label{q2fig}Evolution of the lattice coloring in the two--color case. The area of the lattice is kept at constant 100 unit areas, while the lattice constant $a$ is reduced. The number of sites is shown in the axes.}
\end{figure}

The results of the lattice computations for the two--spin system are shown in Fig.~\ref{q2fig}. The figures show the evolution of the coloring of the $10 \times 10$ unit area lattices, as the lattice constant is reduced from 0.1 to 0.0125. In all cases the configurations indeed correspond to "stripes", which straighten and consist of approximately hexagonal elements, joined by short straight sections, towards smaller $a$. The coarser lattices show clear volume effects, which are reduced in computations with smaller $a$. The value of $\eps(a)$ quickly settles to around 0.29. As noted earlier, the border regions of the lattice have less interacting neighbors inside the lattice, causing the computed $\eps(a)$ to underestimate the true probability. In this case the value of $\eps(a)$ may also vary between lattices simply due to location of the boundaries with respect to the stripes. Visual estimation of the results suggests that the width of the stripes is below 1. None of the sites have exactly zero same--colored neighbors unit distance apart.

\begin{figure}[t!]
	\centering
	\includegraphics[width=0.9\linewidth]{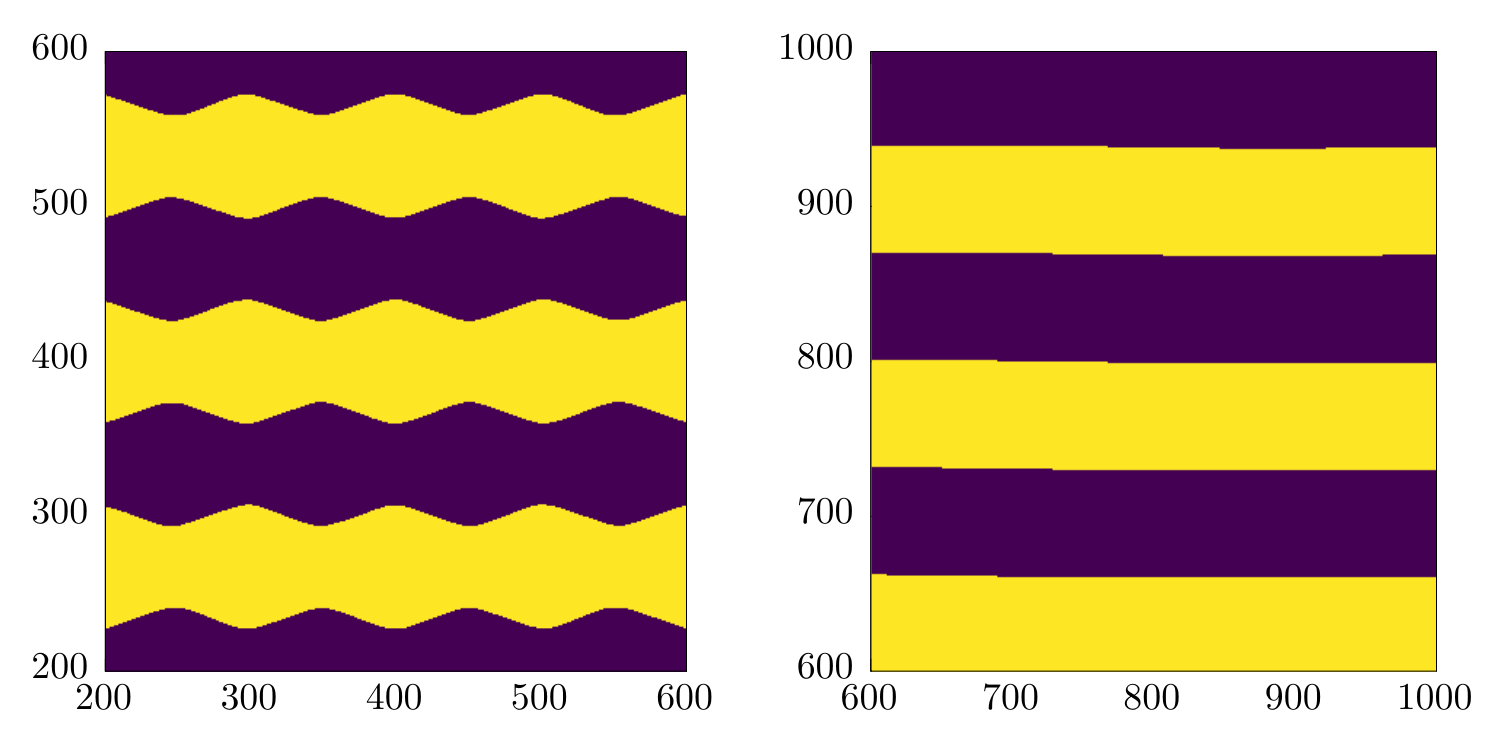}
	\caption{\label{twolat}Computed lattice colorings for a $10 \times 10$ (left) and $20 \times 20$ (right) unit areas with lattice constant $a = 0.0125$. Note that only central portions of lattices are shown.}
\end{figure}

The roughly hexagonal form of the stripes in the $a = 0.0125$ lattice cannot emerge due to its optimal properties, and the shape is likely a volume effect (either due to the boundaries or due to the finite $a$ favoring an otherwise suboptimal configuration). To investigate that further, another computation for the same lattice constant, but this time with a $20 \times 20$ area ($1600 \times 1600$ sites) lattice, was performed. The central parts of the $10 \times 10$ and $20 \times 20$ area lattices are shown in Fig.~\ref{twolat}. The larger lattice has approximately straight stripes covering 69 lattice sites, corresponding to width $69a \approx 0.863$. The result suggests that the hexagonal shape in the smaller lattice probably arises due to the boundaries, and the larger lattice is needed to reveal the true behavior.

At this point we may look how the probability for the striped lattice depends on the width $h$ of the stripes. For comparison I also present the corresponding result for another configuration which easily springs into mind: a checkerboard--colored plane. The checkerboard is taken to consist of squares with side length $h$. Denote the probabilities for the striped plane and the checkerboard--coloring by $p_s(h)$ and $p_c(h)$, respectively. If the stripes are horizontal, one may choose a reference stripe, whose lower edge serves as the origin of the vertical coordinate $z$. Introduce the indicator function
\be
f(z,\phi,h) = 1 - {\rm mod} \left[ {\rm floor}\left(\frac{z + \sin \phi}{h}  \right), 2\right]
\ee
which equals 1 if the point $(\cos \phi, z + \sin \phi )$, unit distance apart from $(0,z)$, has the same color as $(0,z)$, and zero otherwise. Then the probability $p_s(h)$ is computed by averaging over the reference stripe
\be
p_s(h) = \frac{1}{2 \pi h}\int_0^h {\rm d}\,z \int_0^{2\pi}{\rm d}\,\phi \;
f(z,\phi,h).
\label{faves}
\ee
For the checkerboard, the probability $p_c(h)$ is computed similarly. First introduce the indicator function
\be
f(z_1,z_2,\phi,h) = 1 - {\rm mod} \left[ {\rm floor}\left( \frac{z_1 + \cos \phi }{h} \right) + 
{\rm floor} \left( \frac{z_2 + \sin \phi}{h} \right), 2\right],
\ee
where the coordinates $z_1$ and $z_2$ run over a reference square, whose lower--left corner is the origin. Averaging over the reference square gives the desired probability:
\be
   p_c(h) = \frac{1}{2 \pi h^2}\int_0^h {\rm d}\,z_1 \int_0^h {\rm d}\,z_2 \int_0^{2\pi}{\rm d}\,\phi \;
   f(z_1,z_2,\phi,h).
   \label{favec}
\ee
\begin{figure}[t!]
	\centering
	\includegraphics[width=0.8\linewidth]{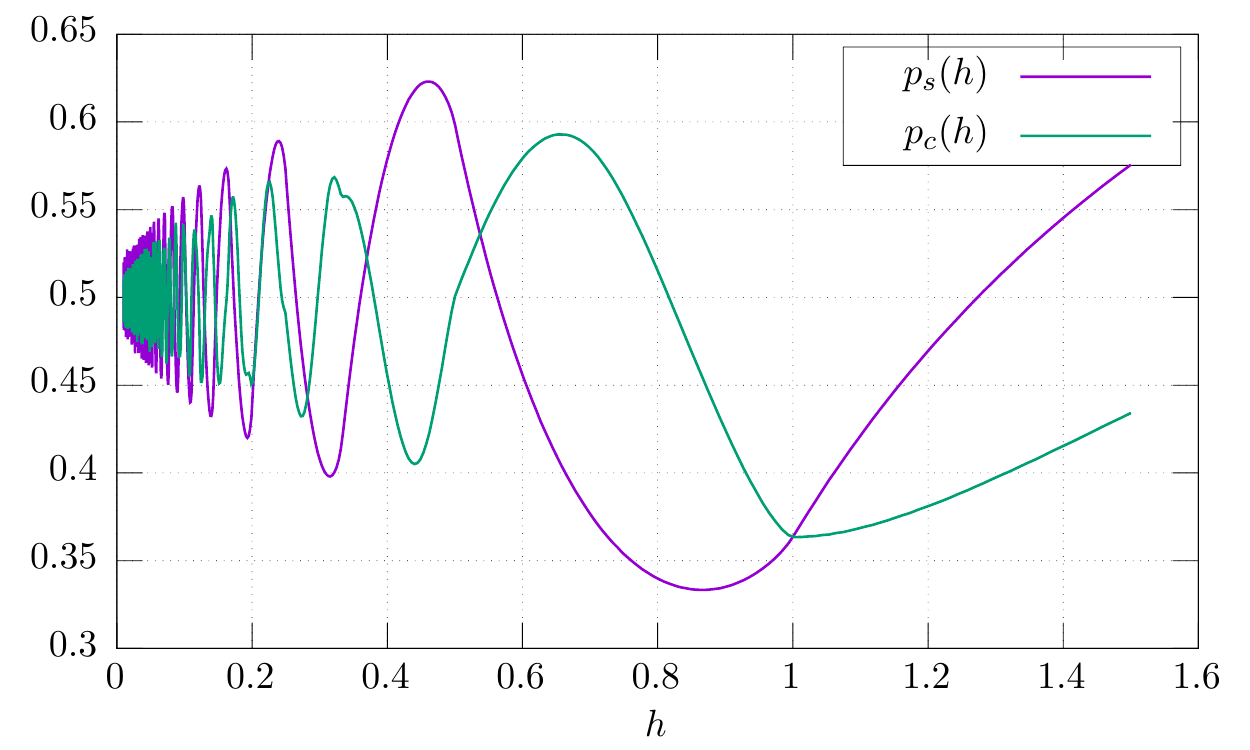}
	\caption{\label{phfig}Probability that two randomly chosen points unit distance apart have the same color on a plane two--colored with alternating straight stripes of width $h$ (curve $p_s(h)$) and with a checkerboard coloring with squares of side length $h$ (curve $p_c(h)$), as a function of $h$.}
\end{figure}
Numerical estimates for  $p_s(h)$ and $p_c(h)$, as a function of $h$, are shown in Fig.~\ref{phfig}. The minima of the probabilities can be inferred approximately from the curves, showing that for stripes the minimum probability is realized in the region $h \geq 1/2$, whereas for the checkerboard the minimum is at $h \geq 1$. In these regions the averagings according to~\eqref{faves} and~\eqref{favec} can be performed analytically, giving
\be
p_s(h) =  1 - \textrm{Re}\left[\frac{4 h \arccos(h) - 4 \sqrt{1-h^2} +2}
{\pi  h} \right], \; h \geq 1/2
\label{pstripes}
\ee
for the striped plane, and
\be
p_c(h) = 1 - \frac{4h-2}{\pi h^2}, \; h \geq 1
\ee
for the checkerboard. The locations of the minima and the corresponding probabilities are accordingly
\setlength{\jot}{-0.5ex}
\begin{gather*}
\begin{aligned}
h &= \frac{\sqrt{3}}{2}, \;
&&p_s = \frac{1}{3} \; &&\text{(optimal straight stripes)}\\
 h &= 1, \;
&&p_c = 1 - \frac{2}{\pi} \approx 0.363 \; &&\text{(optimal checkerboard)}
\end{aligned}
\end{gather*}
confirming that indeed the striped configuration realizes a smaller energy for the two--spin system, corresponding to smaller probability of finding equally--colored points unit distance apart. The result for the optimal width of the stripes $\sqrt{3}/2 \approx 0.866$ reproduces the previously known result~\cite{proba}, and agrees with the simulation result 0.863 for the $20 \times 20$ lattice within the resolution permitted by the lattice constant.

What is important in this comparison is that simulated annealing is seen to reach the correct global minimum among the landscape of solutions exhibiting an infinite amount of local minima on the left--hand side of the global one, as seen in Fig.~\ref{phfig}. The checkerboard behaves in this respect similarly to the striped plane, and while the lowest probability is rather close to the striped plane optimum, those configurations were never realized in the computations. These results, while far from conclusive, give faith that the method is able to reach optimal configurations, even when the underlying set of solutions is very complex.

\begin{figure}[h]
	\centering
	\includegraphics[width=0.8\linewidth]{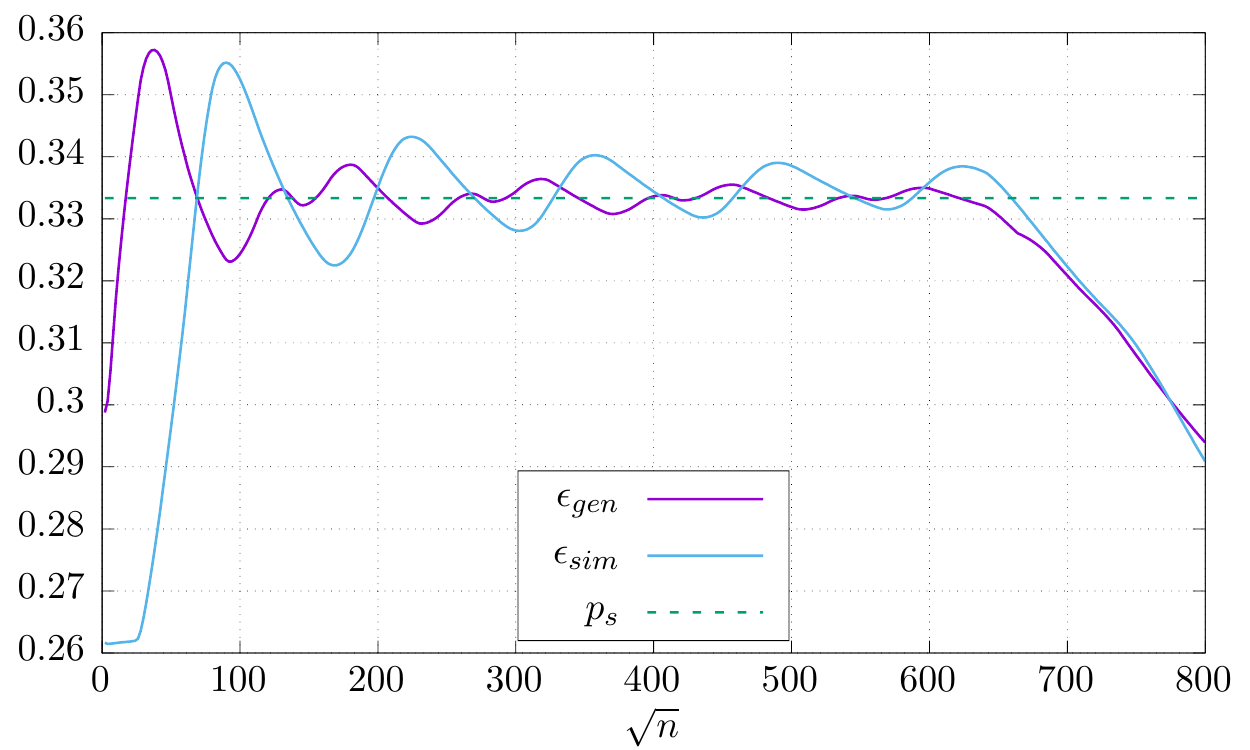}
	\caption{\label{comparefig} The quantities $\eps^{(n)}(a)$ in~\eqref{epsn} for a $10 \times 10$ area ($800 \times 800$ sites) lattice with $a=0.0125$. The curve $\eps_{gen}$ is for the artificially generated lattice of straights stripes, $\eps_{sim}$ shows the probability for the simulated lattice in Fig.~\ref{q2fig}. The theoretical result $p_s$ for the stripe lattice is shown as a dashed line.}
\end{figure}

There is still the question of how closely $\eps(a)$ reproduces the theoretical concept of probability. The computed values for the $10 \times 10$ and $20 \times 20$ lattices were 0.291 and 0.311, respectively. The latter lattice however realizes the optimal probability. To investigate what regions of the lattice contribute to $\eps(a)$ in what way, we use a lattice consisting of straight stripes as a test. We artificially construct an $10 \times 10$ area lattice with $a=0.0125$, having straight stripes of width $a n_s$, where $n_s = \textrm{int} \left[ \sqrt{3}/2a \right] = 69$ is the number of lattice sites covered by the stripe (int: nearest integer value). We then compare $\eps(a)$ of this lattice (denoted as $\eps_{gen}$) to the corresponding simulated lattice ($\eps_{sim}$) in Fig.~\ref{q2fig}. According to~\eqref{ene}, $\eps(a)$ represents the average over the whole lattice. If we compute the average cumulatively, starting from the central part of the lattice, we can plot the quantity
\be
   \eps^{(n)}(a) = \frac{1}{n} \sum_{k=1}^{n} \eps_k (a)
   \label{epsn}
\ee
for the central $n$ site sublattices as $n=1, \ldots ,N$. This gives a way to estimate how the boundary regions of the lattice affects the value, and how we can get a reliable estimate for the theoretical probability from the computations. Fig.~\ref{comparefig} shows the probabilities computed according to~\eqref{ene} for the artificially generated and the simulated lattices. The figure also shows, as a straight line, the theoretical probability according to~\eqref{pstripes}. The results show that the small value of $\eps(a)$ for the simulated $10 \times 10$ area lattice is indeed due to the boundary region, as $\eps(a)$ drops below the corresponding straight stripe lattice value only just before the boundary is reached. The inner parts the straight stripes lattice show comparable or even smaller probability. The figure also shows that the value of $\eps(a)$ is indeed similar to the theoretical probability in~\eqref{pstripes} and that the finite lattice of $10 \times 10$ unit areas is large enough to realize this connection. This requires that we consider the portion of the lattice, where all spins have all interacting neighbors inside the lattice. 
There is no obvious best choice for comparing the value $\eps(a)$ and the theoretical concept of the probability, but in what follows we estimate the ``correct'' values as being approximately where the cumulative average as defined by~\eqref{epsn} has reached a plateau, similar to what is seen in Fig.~\ref{comparefig}. For larger number of colors below, the behavior of the cumulative average is more difficult to analyze though.

\subsection{Three and four colors}

\begin{figure}[t!]
	\centering
	\includegraphics[width=1.0\linewidth]{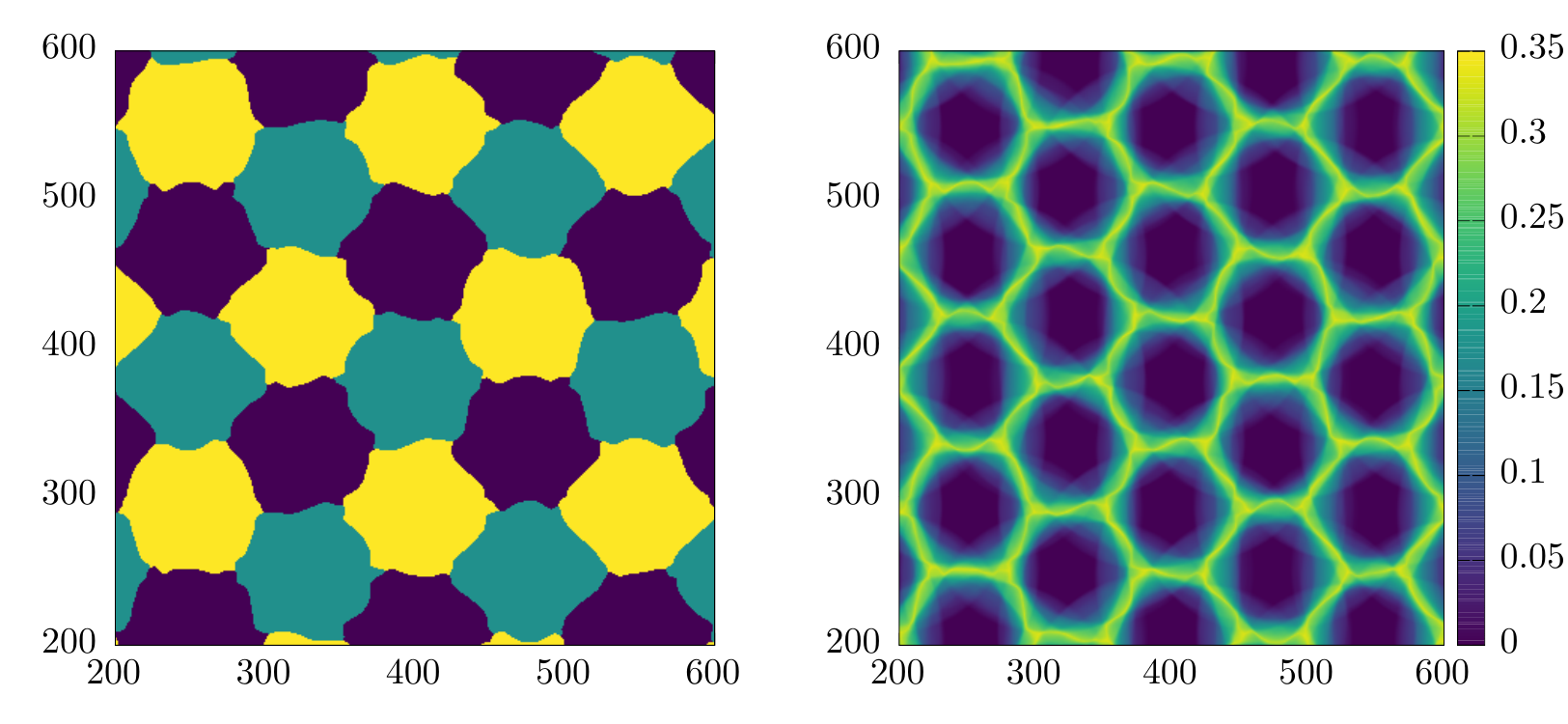}
	\caption{\label{cp_q3} Computed optimal three--coloring of the lattice (left) for lattice constant $a=0.0125$, and the associated heat map showing the distribution of the values $\eps_k (a)$ defined in~\eqref{epsk}, corresponding to the fraction of neighbors with same color unit distance apart (right). For clarity,  only the central 25 unit areas of the $800 \times 800$ site lattices are shown.}
\end{figure}

 The lattice simulation for a three colors is shown in Fig.~\ref{cp_q3}. The right pane of the figure is a heat map containing the values $\eps_k (a)$ in~\eqref{epsk} over the lattice. It shows with color coding the locations of spins having high energies, which correspond to those spins possessing same--valued neighbors unit distance apart. This kind of map of "non--colorability" is useful in examining the nature of the solutions. It seems natural to take regular (periodic) frustrations to describe a property of the solution, effects which are not due to numerical errors or noise, but originating from the fact that the lattice spin system has reached at least a local minimum where there is an obstacle for further lowering of energy. 

The computed optimal three--coloring map has an approximate hexagonal symmetry, and consists of distorted hexagons as individual colored areas. The structure is also visible in the heat map, showing that the frustrations (spins with positive energy) are arranged hexagonally, albeit with wiggly boundaries. The values of $\eps_k (a)$ get gradually smaller towards the centers of the hexagons. Seventeen per cent of the lattice area is colored, i.e.~has no same--colored sites unit distance apart. It is perhaps interesting that in the calculations with larger lattice constants the hexagonal structure of the lattice was more regular, and the distorted boundaries only emerged around $a=0.025$, and were even more irregular for the displayed lattice. This was however accompanied with slowly decreasing $\eps(a)$, so it might be a real effect. 

The value of $\eps(a)$ is best inferred in a method similar to Fig.~\ref{comparefig}, as a cumulative average. For the three--colored lattice the best computations here give $\eps \approx 0.12$, which conforms with the analytical lower bound of $1/11$~\cite{proba} for the three--coloring.

\begin{figure}[t!]
	\centering
	\includegraphics[width=1.0\linewidth]{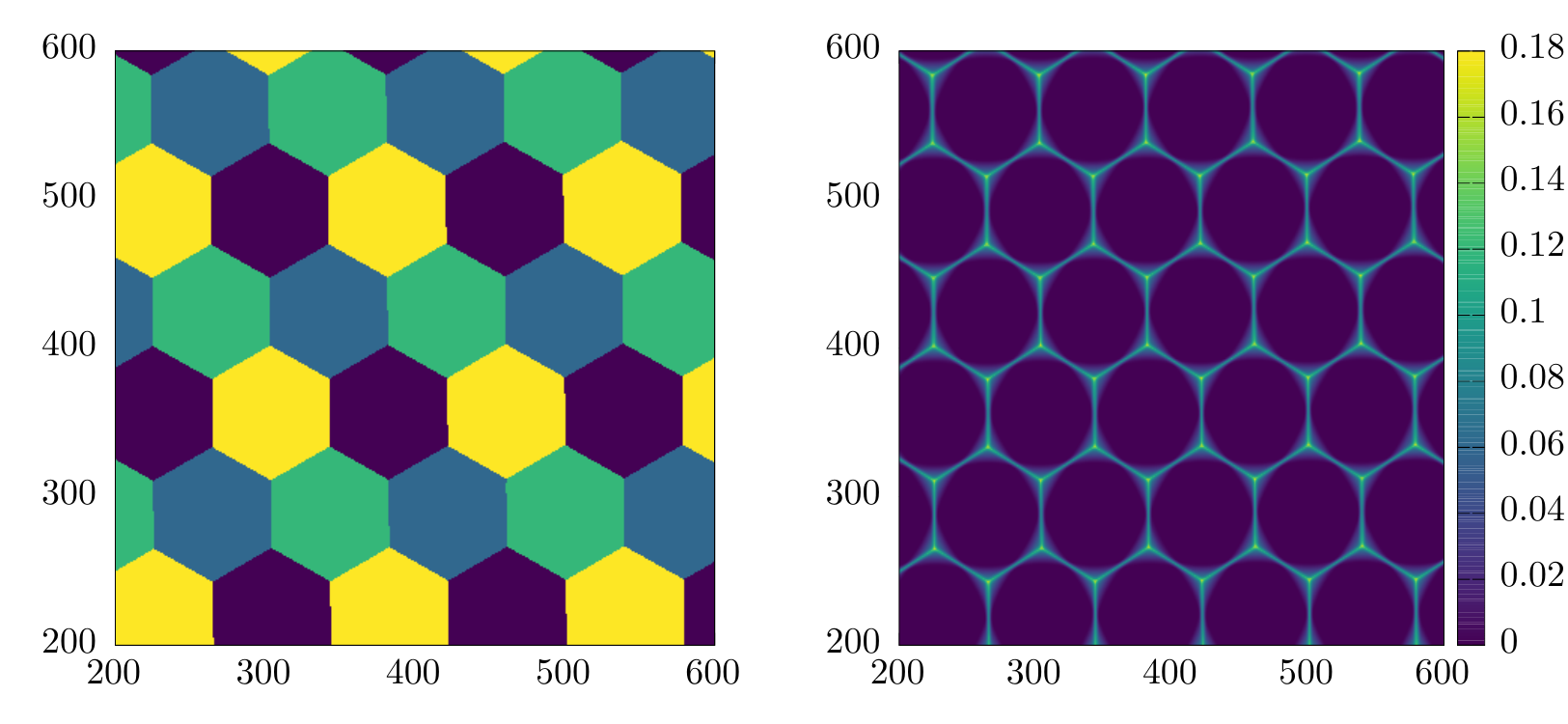}
	\caption{\label{cp_q4} Computed optimal four--coloring of the lattice (left), and the associated heat map showing the fraction of neighbors with same color unit distance apart (right). The lattice had the size of $10 \times 10$ unit areas and the lattice constant was $a=0.0125$. The central 25 unit areas of the lattices are shown.}
\end{figure}

The four color simulation results also in a hexagonal coloring, as shown in the left hand side Fig.~\ref{cp_q4}. The converged value for the relative energy $\eps(a)$ was $\sim 0.01$, an order--of--magnitude improvement over the three--color case.
The individual colored areas are this time regular hexagons, which themselves form a hexagonal lattice with the usual basis vectors
\be
  \mathbf{u}_1 = \alpha \, \mathbf{i}, \; \; 
  \mathbf{u}_2 = \frac{\alpha}{2}\, \mathbf{i} + 
  \frac{\sqrt{3} \alpha}{2}\, \mathbf{j},
\ee
where the unit of the hexagonal lattice $\alpha = \sqrt{3}s$, in terms of the side length $s$ of the individual hexagons. Hexagons forming a monocolor sublattice are described by lattice vectors $2n\mathbf{u}_1 + 2m\mathbf{u}_2$, with integer $n, m$. Based on the simulation result with the $a=0.0125$ lattice constant, an individual hexagon covers 79 lattice sites in the horizontal direction, corresponding to $\alpha \approx 0.988$, just below one (at this resolution, 80 sites would already cover a unit distance). I was not able to analytically estimate an optimal value for $\alpha$, using the method that worked for the two--color case. However, some other method might well do the trick and it would be interesting to see if the numerical result is indeed optimal for the regular hexagonal configuration.

The distribution of the values $\eps_k(a)$ is shown in the right pane of Fig.~\ref{cp_q4}. The structure is correspondingly regularly hexagonal, with the highest values around 0.18 being present in the vertices, where they appear as bright spots in the lattice. The majority of lattice sites has $\eps_k(a) = 0$. The colored area has reached 72\% of the whole lattice.

\subsection{Five and six colors}

\begin{figure}[t!]
	\centering
	\includegraphics[width=1.0\linewidth]{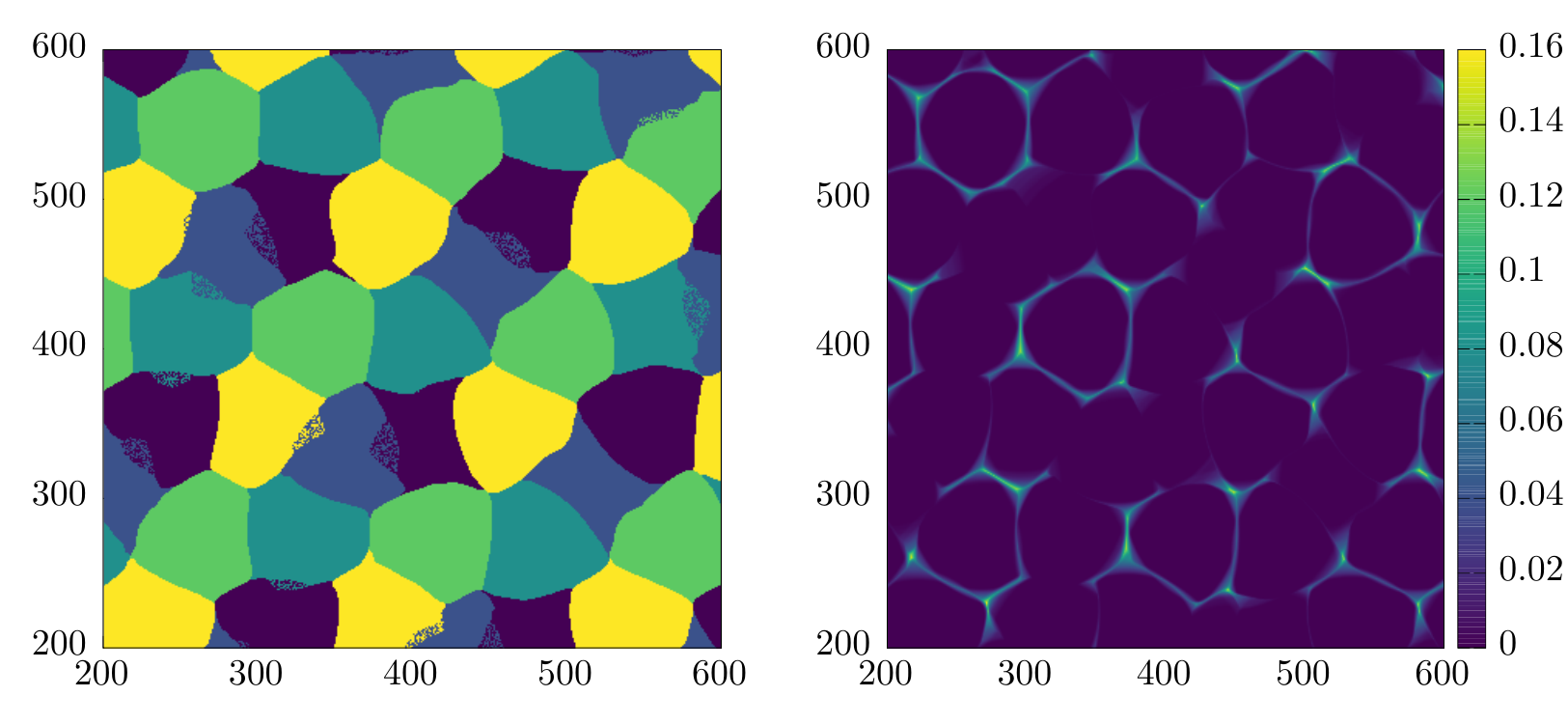}
	\caption{\label{cp_q5} Computed optimal five--coloring of the lattice (left), and the associated heat map showing the fraction of neighbors with same color unit distance apart (right). The lattice used in the simulation had the size of $10 \times 10$ unit areas and lattice constant $a=0.0125$. The central 25 unit areas of the lattices are shown.}
\end{figure}

Unlike the map--like colorings above, the computed optimal five--coloring of the lattice in Fig.~\ref{cp_q5} includes diffuse areas, indicating locations where spins can have either of the two values without affecting the energy. The lattice coloring has an approximate hexagonal symmetry. The heat map of the distribution of $\eps_k (a)$ values exhibits a quasi-regular structure showing that the border regions of the colored areas have frustrations preventing further lowering of energy. The value of $\eps(a)$, estimated as in Sec.~\ref{kakkonen}, is roughly $\sim 4 \times 10^{-3}$. This is again an order--of--magnitude drop from the previous case. The largest individual values of $\eps_k (a)$ were around 0.16, the same order as in the four--color case. This might be partly due to the poor quality of the five--color computation, as the lattice of $\eps_k (a)$ values is not very regular. Again the highest energy spins concentrate to form bright spots in the lattice.
The computation shown in Fig.~\ref{cp_q5} used the lattice constant $a=0.0125$, and the results were not essentially improved by lowering the value. The computations with smaller $a$ used smaller lattices, but this was hardly the reason for the stalling of convergence, as the structures of both the spin lattice and $\eps_k(a)$ values were quite similar in shape. The colored area covers 81\% of the lattice, which is not a dramatic improvement over the previous case of four colors, again perhaps a sign that the quality of the simulation was not very high.


There are reasons to believe that the case of six colors can be difficult to resolve with this method. This is due to the existence of explicit "almost six--colorings": seven colorings with a small amount of the seventh color. An example constructed by Soifer~\cite{soifer} has a ratio $\sim 1/300$ for the areas of the seventh color to the others. It is possible that the annealing method could converge towards the correct minimum, but be unable to tell the difference between zero and finite but very small $\eps(a)$. In computations the value $\eps(a)$ was seen to decrease for decreasing $a$, until a plateau was reached around $a=0.01$. The smallest values obtained for $\eps(a)$ were $\sim 2 \times 10^{-4}$. This is hardly the absolute minimum value possible with these lattice constants, but probably shows the correct order of magnitude. A more reliable estimate could likely be given using larger lattices.

\begin{figure}[h!]
	\centering
	\includegraphics[width=1.0\linewidth]{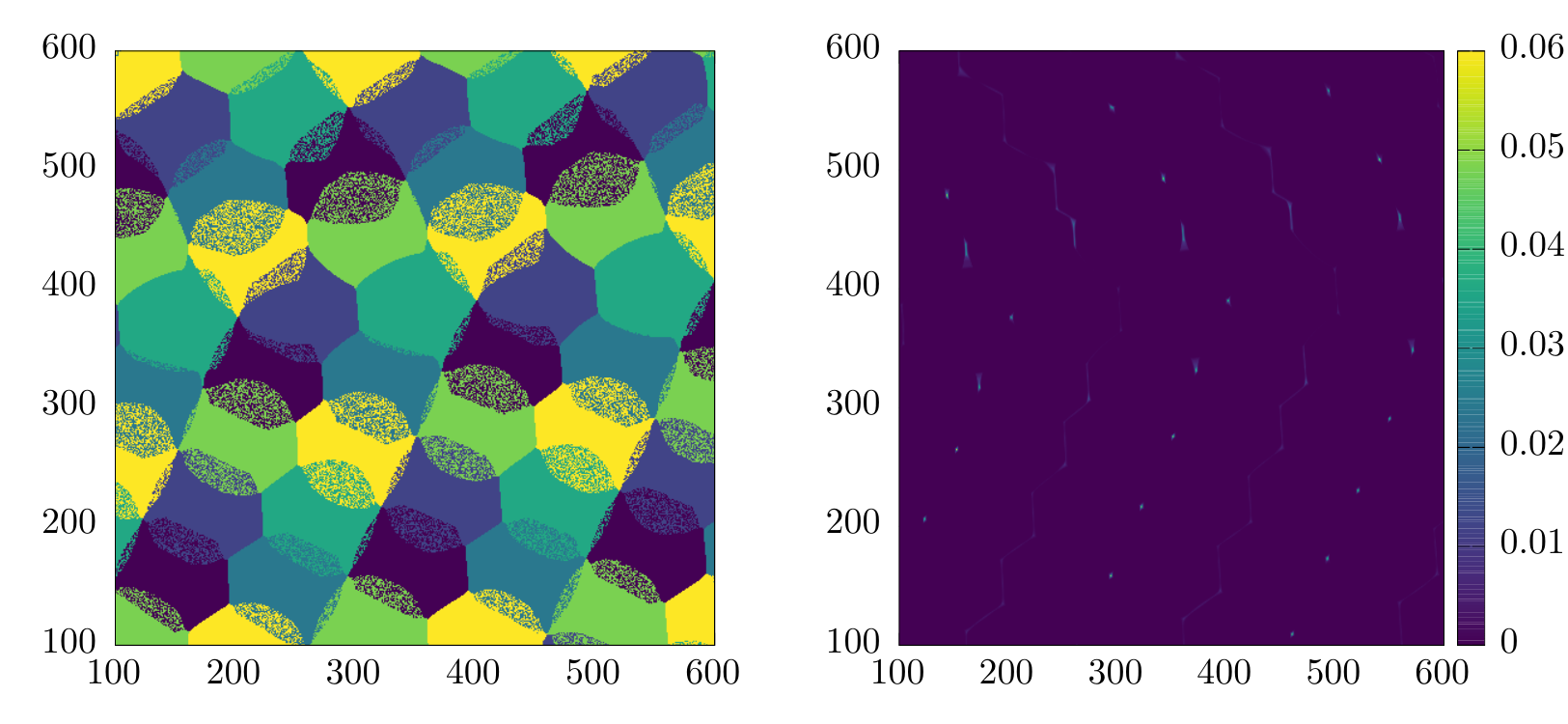}
	\caption{\label{cp_q6} Computed optimal six--coloring of the lattice (left) for lattice constant $a=0.01$, and the associated heat map showing the fraction of neighbors with same color unit distance apart (right). The central 25 unit areas of the lattices are shown.}
\end{figure}

The distribution of frustrations in the lattice for the all the best converged calculations with six spins formed regular structures, as seen in the example calculation in Fig.~\ref{cp_q6}. This figure pair shows the central portions of a $700 \times 700$ lattice and the associated heat map, computed using the lattice constant $a=0.01$. 
The heat map may not be easy to inspect visually, but the distribution of values of $\eps_k(a)$ consists of lines of point--like structures, with wiggly lines in between. This particular computation shows a wedge in the middle of the lattice, where the direction of the structures changes. This is most likely a volume effect, induced by border regions. The lattice structure formed by frustrations $\eps_k (a)$ were similar in other converged six--color computations. This phenomenon can be contrasted to the seven color case discussed below, where even in the calculations not converging to zero the frustrations were irregular domains, giving the impression of numerical noise. The colored area is now 97\% of the whole lattice. More comprehensive computations, both using smaller lattice constants and larger lattices, should be able to confirm is this is a real effect. This would likely mean a decisive obstacle preventing a six--coloring, and pointing to the chromatic number of the plane equaling seven. 

\subsection{Seven colors}

\begin{figure}[t!]
	\centering
	\includegraphics[width=0.7\linewidth]{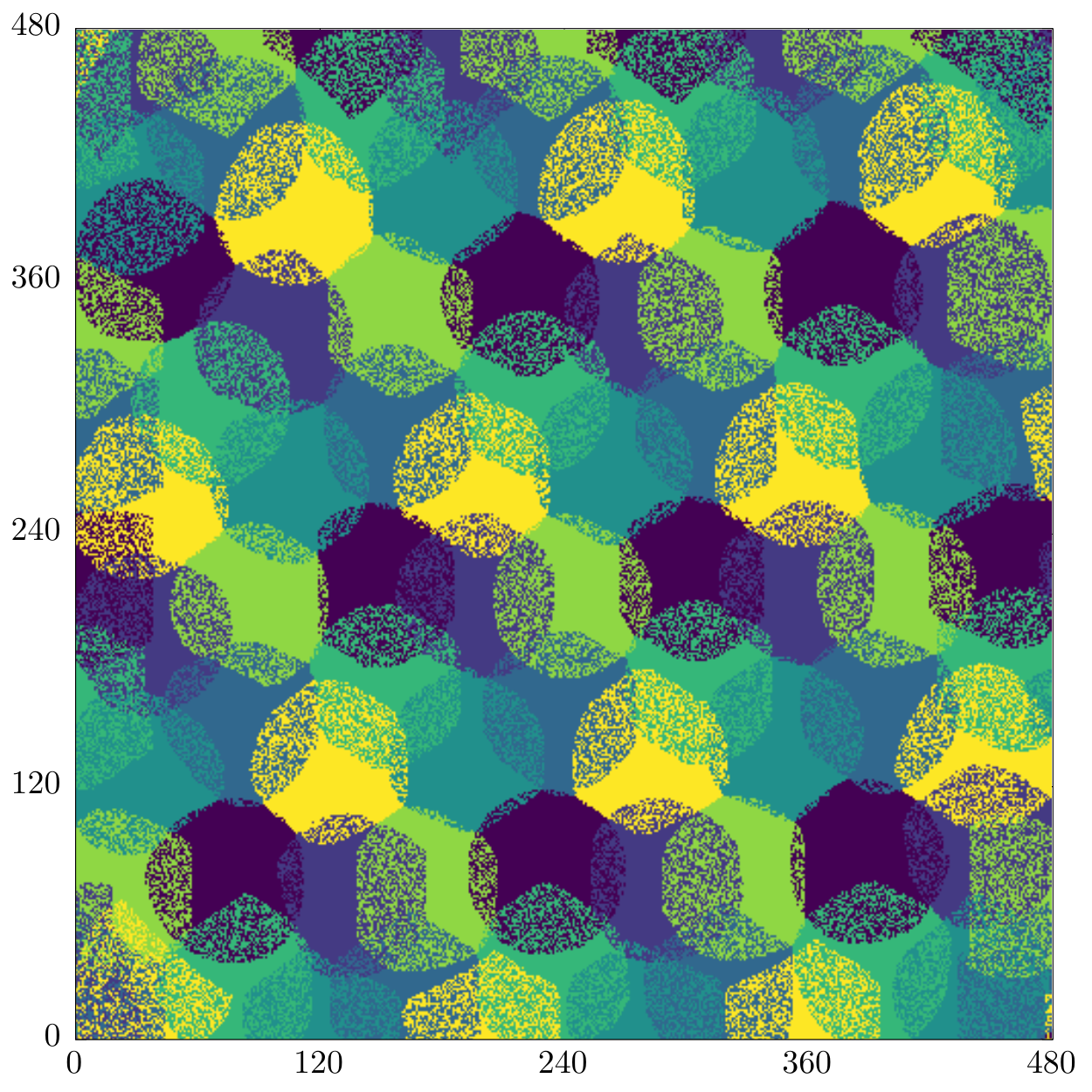}
	\caption{\label{cp_q7} Computed optimal seven--coloring of the lattice for a lattice constant $a=0.0125$. The whole lattice ($6 \times 6$ unit areas) is shown. The computation converged to zero energy within the precision used in the computations. This means that every pair of spins unit distance apart are different.}
\end{figure}

The case $q=7$ is important, as it is known through explicit constructions that the plane is seven--colorable. Our method should therefore, if it works as advertised, converge towards zero energy for seven and higher colors. This indeed happened in the computations. A typical computation did not converge to zero, but a successful example is shown in Fig.~\ref{cp_q7}. What is important is that this computation was done with a relatively modest lattice having $480 \times 480$ sites and lattice constant $a=0.0125$, corresponding to lattice size of $6 \times 6$ unit areas. The simulation used 110 steps in temperature, with the scaling factor 0.83 between temperature steps, and for each temperature value the lattice was updated 300 times. The values $\eps_k (a)$ in~\eqref{epsk} all converged to zero within the double precision arithmetic. This amounts to all lattice sites unit distance apart having a different spin, and thus a coloring of the lattice was achieved. This shows that the lattice constant does not need to be very small for convergence to zero to happen, or that volume effects do not necessarily spoil the computation for small lattices. This is one of the reasons (besides computational cost) that the previous cases $q=5,6$ were not done with very much smaller lattice constants. 
There are lot of diffuse domains visible in the lattice, meaning that there is apparent freedom in choosing the colors in these areas. There is some resemblance in the picture to the known explicit construction (see e.g.~Soifer's book~\cite{soifer}), which consists of hexagonal elements. The shapes of individual colored areas are not very regular, which points to lot of freedom and probably a continuum of seven--colorings. Since the coloring is not strictly periodic, this particular result cannot be generalized to obtain a coloring of the infinite lattice. One is tempted to believe however that the above result is not an artifact of the finite lattice and that the there is a corresponding coloring for the infinite lattice also.

There is more than one exact seven--coloring of the plane~\cite{soifer}. This means that the ground state of the present model must be degenerate for seven spin values and above, at least in the continuum limit. Thus it is conceivable that the solution found by simulated annealing might interpolate between these ground states in the lattice, as they would appear locally equally attractive to the Monte Carlo search. This could be the reason why in most calculations with seven colors there were random small frustrations present in the lattice.

\section{Conclusions}

Lattice simulations of the fixed--range interaction multicomponent spin model introduced here produce approximations to optimal plane colorings as ground states of the system. These optimal colorings realize a minimum probability for points unit distance apart in the plane having the same color. The simulations are in agreement with essential previous knowledge about plane colorings: the optimal two--coloring consists of straight stripes, corresponding to probability $1/3$, and in general the probability is nonzero for less than five colors. On the other hand for seven colors a convergence towards zero probability was seen. For the open cases of five and six colors definite conclusions could not be given based on simulations performed here, but the results were suggestive: for five and six colors the simulations leading to smallest probabilities were not essentially improved by lowering the lattice constant below 0.01. For all cases with less than seven colors the lattice colorings showed approximately periodic frustrations. These were areas where the spins had the highest energies, corresponding to equal colors unit distance apart. The regularity of these areas points to systematic sources of non--colorability, which are not due to numerical errors. The other possibility for five and/or six colors not converging to zero would be that simulated annealing could not find the true minima in the energy landscape of the spin system. This is an inherent limitation of the method, preventing absolute conclusions and meaning that the results provide guidance but not proofs.

The reasons for potentially ending up in the wrong minimum include volume effects, collectively describing phenomena emerging due to the final size of the lattice. For example the unavoidable border regions in finite lattices could induce configurations having almost similar but not quite the same energies as those corresponding to a global minimum, and which propagate further and prevent finding the true optima of the ideal infinite lattice. An example of this was probably present in the two--color case discussed in Sec.~\ref{kakkonen}. On the other hand, volume effects were not always fatal, as is evident from the converged seven--color computation shown in Fig.~\ref{cp_q7}, where a rather modest lattice of $6 \times 6$ unit areas was used. The Hamiltonian that the model builds on has translational symmetry, and the resulting colorings should reflect this property at long distances. Therefore it should be possible to obtain reliable results with finite (and not very large) lattices. For the two--color case, comparison of the simulation result and an analytical solution showed that annealing found the true global minimum (the striped configuration), even though an infinite amount of local minima were available. In addition, checkerboard configurations with rather similar energies were systematically absent in the results. 

Considering the implementation of the simulations, there are many options not investigated here. These include other possible forms of the distribution function used to approximate the delta function in the lattice. Also, only square lattices were used in this work, but for example hexagonal lattices might bring some advantages. There are also numerous possibilities considering the implementation of the simulated annealing algorithm, most importantly the rules employed to update the spins. Finding out what effect, if any, these other options have on the convergence properties of the simulations could add valuable insight.

The method and results presented here naturally give rise to the question of optimal colorings of the plane with few colors. Specifically:
\begin{enumerate}
	\item Are the simulation results for three and four color lattices optimal? Especially the regular hexagonal lattice in the four--color case is interesting. Simulation result gave the unit of the lattice just below one, and there is probably an analytical derivation giving the exact value for lattice unit, and especially the probability, which was numerically estimated to be roughly 0.01.
	\item Are there true five or six colorings of the plane, or are the apparent convergence of energy and the regular frustrations in lattice calculations a sign of fundamental non--colorability? More extensive simulations could point to correct answers. Here convergence to zero energy (probability) was only seen for seven colors, together with the absence of the periodic frustrations even for poorly converged calculations. This points to the chromatic number of the plane equaling seven.
        \item Could the energy functional $\eps(a)$ in~\eqref{ene} be investigated by other methods than the numerical approach employed here? For example, could the simulated optimal configurations be used as a starting point for further analysis?
\end{enumerate}

In addition, the method presented here can be generalized or applied to different situations. Some possibilities are listed here:
\begin{enumerate}
	\item By changing the boundary conditions of the lattice spins, other topologies can be studied. For example, periodic boundary conditions in one dimension give approximations to cylinder colorings, periodicity in two dimensions to colorings on a torus. 
	\item The method can in principle be applied to the chromatic number in higher dimensions, but the obvious downside is the large computational cost associated.
	\item Other coloring problems can be studied by modifying the Hamiltonian. For example, multiple distances could be avoided by adding the appropriate interaction terms, or different colors can be made to avoid different distances. The second case refers to the \textit{polychromatic number of the plane}~\cite{soifer}, as defined by Erd\H{o}s.
\end{enumerate}

I hope this paper has been of interest to friends of recreational and experimental mathematics, and stimulates readers to further tests and investigations. I am grateful to Mikko Laine for discussions and comments.

\bibliography{colors}{}

\end{document}